\documentclass[sigconf]{acmart}

\usepackage{amsfonts}       
\usepackage{amsmath}
\usepackage{graphicx, subfig}
\usepackage{nicefrac} 
\usepackage[ruled, vlined, linesnumbered]{algorithm2e}
\usepackage{enumerate}
\usepackage{enumitem}
\usepackage{geometry}
\usepackage{tikz}
\usetikzlibrary{shapes.geometric}
\usepackage{arydshln}
\usepackage{xspace}

\newtheorem{definition}{Definition}

\def\header{\vspace{2mm} \noindent}

\AtBeginDocument{%
  \providecommand\BibTeX{{%
    \normalfont B\kern-0.5em{\scshape i\kern-0.25em b}\kern-0.8em\TeX}}}

\copyrightyear{2022}
\acmYear{2022}
\setcopyright{acmlicensed}\acmConference[KDD '22]{Proceedings of the 28th ACM SIGKDD Conference on Knowledge Discovery and Data Mining}{August 14--18, 2022}{Washington, DC, USA}
\acmBooktitle{Proceedings of the 28th ACM SIGKDD Conference on Knowledge Discovery and Data Mining (KDD '22), August 14--18, 2022, Washington, DC, USA}
\acmPrice{15.00}
\acmDOI{10.1145/3534678.3539390}
\acmISBN{978-1-4503-9385-0/22/08}

\begin{document}
\title{Sampling-based Estimation of the Number of Distinct Values in Distributed Environment}

\author{Jiajun Li}
\affiliation{%
  \institution{Renmin University of China}
}
\email{2015201613@ruc.edu.cn}

\author{Zhewei Wei}
\authornote{Zhewei Wei is the corresponding author. The work was partially done at Gaoling School of Artificial Intelligence, Peng Cheng Laboratory, Beijing Key Laboratory of Big Data Management and Analysis Methods and MOE Key Lab of Data Engineering and Knowledge Engineering.}
\affiliation{%
  \institution{Renmin University of China}
}
\email{zhewei@ruc.edu.cn}

\author{Bolin Ding}
\affiliation{%
  \institution{Alibaba Group}
}
\email{bolin.ding@alibaba-inc.com}

\author{Xiening Dai}
\affiliation{%
  \institution{Alibaba Group}
}
\email{xndai@live.com}

\author{Lu Lu}
\affiliation{%
  \institution{Alibaba Group}
}
\email{lu.lu@alibaba-inc.com}

\author{Jingren Zhou}
\affiliation{%
  \institution{Alibaba Group}
}
\email{jingren.zhou@alibaba-inc.com}

\renewcommand{\shortauthors}{Jiajun Li et al.}

\begin{abstract}
In data mining, estimating the number of distinct values (NDV) is a fundamental problem with various applications. Existing methods for estimating NDV can be broadly classified into two categories: i) scanning-based methods, which scan the entire data and maintain a sketch to approximate NDV; and ii) sampling-based methods, which estimate NDV using sampling data rather than accessing the entire data warehouse. Scanning-based methods achieve a lower approximation error at the cost of higher I/O and more time. Sampling-based estimation is preferable in applications with a large data volume and a permissible error restriction due to its higher scalability. However, while the sampling-based method is more effective on a single machine, it is less practical in a distributed environment with massive data volumes. For obtaining the final NDV estimators, the entire sample must be transferred throughout the distributed system, incurring a prohibitive communication cost when the sample rate is significant. This paper proposes a novel sketch-based distributed method that achieves sub-linear communication costs for distributed sampling-based NDV estimation under mild assumptions. Our method leverages a sketch-based algorithm to estimate the sample's {\em frequency of frequency} in the {\em distributed streaming model}, which is compatible with most classical sampling-based NDV estimators. Additionally, we provide theoretical evidence for our method's ability to minimize communication costs in the worst-case scenario. Extensive experiments show that our method saves orders of magnitude in communication costs compared to existing sampling- and sketch-based methods.


\end{abstract}

\ccsdesc[500]{Computing methodologies~MapReduce algorithms}

\keywords{sampling, distributed environment, NDV, communication}

\maketitle


\section{Introduction}
Estimating the number of distinct values (NDV) in data mining is a fundamental but critical problem. NDV estimation can improve the efficiency of database tasks~\cite{charikar2000towards,haas1995sampling,ozsoyoglu1991estimating,hou1989processing,naughton1990estimating}, and it also has many applications in other areas. These studies include estimating the unseen species in ecological studies~\cite{bunge1993estimating,valiant2013estimating}, data compression~\cite{lemire2011reordering}, network security~\cite{cohen2019cardinality}, and statistics~\cite{hou1988statistical}. A standard method of calculating NDV involves scanning the table, followed by sorting or hashing. When calculating the exact NDV for any distribution, which takes $O(N\log(N))$ time, where $N$ is the size of the data, sorting is unavoidable. With the help of the streaming algorithms such as the FM Sketch~\cite{flajolet1983probabilistic} and the HyperLogLog Sketch~\cite{flajolet2007hyperloglog}, it is sufficient to scan the table once to provide a high-precision estimation for distinct values. However, scanning the entire table incurs high I/O and time costs as the data grows in size. Therefore, sampling is a frequently used approach for obtaining an acceptable approximate solution in applications where scalability is the primary concern.

\header{\bf Motivation. }The motivation comes from estimating the NDV in large-scale distributed environments. First of all, scanning-based methods incur high I/O costs, limiting their scalability. On the other hand, while sampling-based NDV estimators reduce the I/O cost on a single machine~\cite{brutlag2002block}, they may lead to high communication costs while transmitting the samples. For example, we consider the GEE~\cite{charikar2000towards} estimator. GEE uses the equation $\hat{D}_{GEE}=\sqrt{\frac{1}{q}}f_1+\sum_{i\geq 2}f_i$ to estimate the NDV of data, where $q$ is the sample rate and $f_i$ is the {\em frequency of frequency} of the sample (i.e. $f_i$ is the number of elements that appear exactly $i$ times in the sample). To compute the {\em frequency of frequency} of the sample, we need the frequency dictionaries of items on each machine. Most items are different, which means that the total size of frequency dictionaries is close to the sample size. ~\cite{charikar2000towards} provides theoretical evidence that an adequate sample rate ($\ge 0.1\%$) is needed to obtain a reasonable NDV estimator. Therefore, if the data set has a large number of distinct values, we will end up with a massive dictionary to store the frequency of the samples. If the raw data is at the terabyte level, the sample is at the gigabyte level. In distributed systems where we need to transfer the dictionaries of the samples across the machines, such large dictionaries will lead to high communication costs, limiting the system's scalability. 

\begin{figure}[t]
	\centering
	\includegraphics[width=60mm]{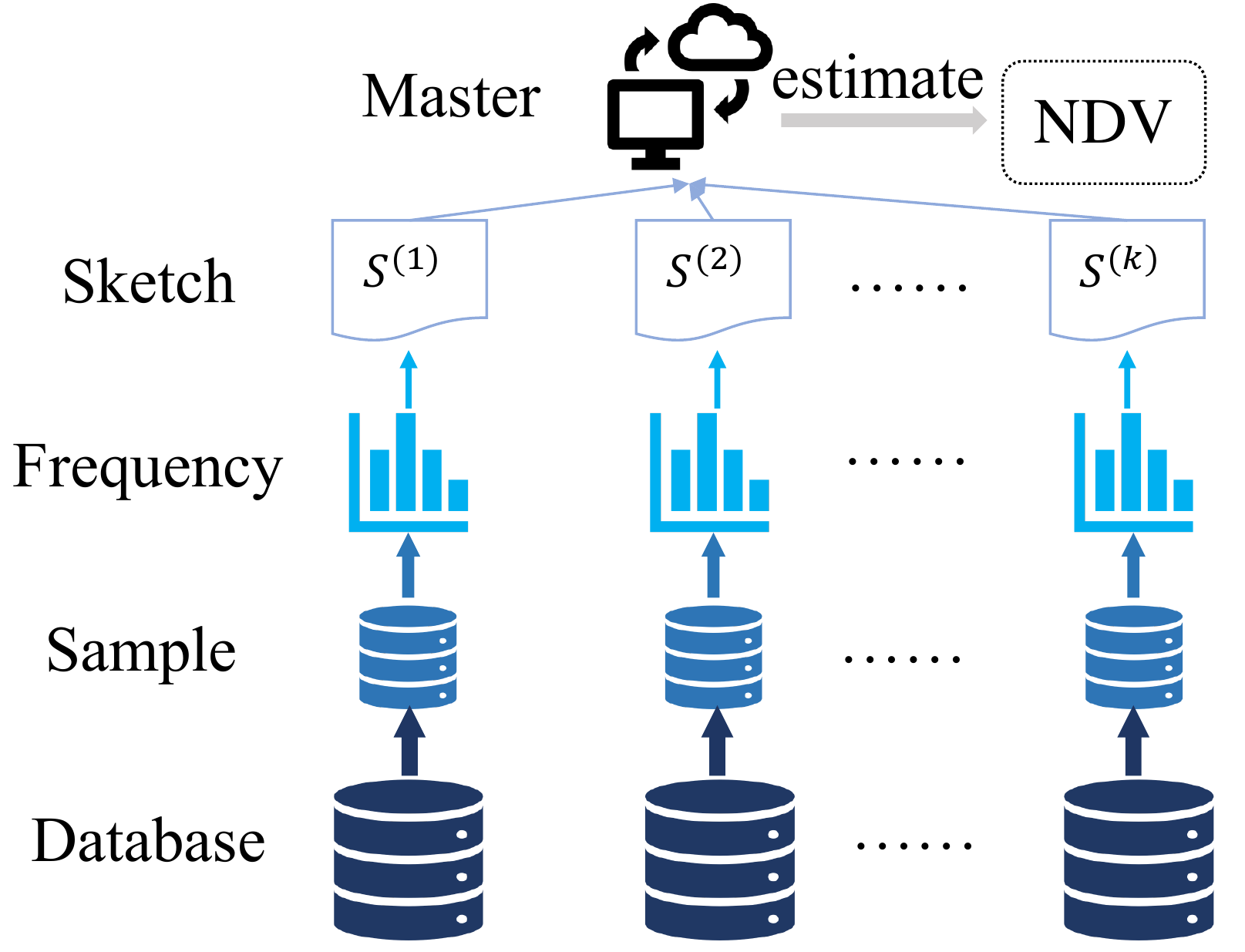}
    \vspace{-1em}
	\caption{Sampling-based  estimation of NDV model in the distributed environment}
	\vspace{-2em}
	\label{fig:model}
\end{figure}
 
\header{\bf Distributed Model for Sampling-based NDV Estimation.} In this paper, we consider the problem of extending various classic sampling-based NDV estimators~\cite{ozsoyoglu1991estimating,shlosser1981estimation,chao1992estimating,charikar2000towards} to distributed settings with sub-linear communication costs.
We assume the target data is partitioned and stored on multiple machines. Rather than directly computing the entire sampling distribution, we provide a novel sketch-based algorithm to avoid transferring the frequency dictionaries of the samples. Additionally, we will provide theoretical demonstrations and experiments to prove that our method will maintain high accuracy while reducing transmission costs. 


Our computation framework is shown in Figure~\ref{fig:model}. We assume that the data is stored in a file system and uniformly distributed across multiple machines. Each machine may store terabytes of data, and the objective is to compute the NDV over the data union. One approach is to scan each database using sketching algorithms such as HyperLogLog\cite{flajolet2007hyperloglog} and then merge the sketches on a master node to avoid high communication costs. However, this approach necessitates scanning the entire database, which results in a high I/O cost. Another possibility is to collect a sample from each machine and merge their dictionaries based on their frequency distributions in the master node. If data is not uniformly distributed, ~\cite{brutlag2002block} also suggests using block sampling to avoid scanning the whole table. While sampling reduces I/O costs, the frequency dictionaries may be significant due to the large data volume and high sampling rate, resulting in a high communication cost. The purpose of this paper is to develop a method that combines the best of both worlds: a sampling-based NDV estimation algorithm that does not require a cross-machine transfer of sample frequencies.

At first glance, the above goal can be achievable by scanning each sample with a HyperLogLog sketch and merging the sketches in the master node. However, this approach yields an estimator for the NDV of the union of samples, not the raw data. Sampling-based NDV estimators, such as GEE, require the sample's $f_1$, which is difficult to estimate with HyperLogLog sketches. As a result, the fundamental challenge is estimating the $f_1$ of the samples without transferring the sample dictionaries between machines.



\header{\bf Our contributions.} This paper estimates the NDV with low communication costs in the distributed environment and makes the following contributions. 
\begin{itemize}[leftmargin = *]
	\item We prove a lower bound that states, in the worst case, it is impossible to estimate the sample's $f_i$ in the distributed environment unless a linear communication cost is allowed. 
	\item We propose a distributed streaming algorithm that achieves sub-linear communication cost for computing the unique values of the sample(i.e., the unique values is the number of elements that appear once), under a mild assumption on the distribution.
	\item We show how our framework can incorporate classic NDV estimators, including GEE, AE, Shlosser, CL1. We also propose several modified estimators to better cope with our framework. 
    \item We conduct extensive experimental studies to demonstrate the scalability of our method. The proposed methods reduce the communication costs by orders of magnitude.
\end{itemize}
\section{Preliminaries}
This section will give a brief review of the related concepts. Table~\ref{tab:notation} summarizes the notations frequently used throughout the remainder of the work.




\begin{table}[t]
	\begin{small}
		\centering
		\caption{Table of notations.}
		\vspace{-1em}
		\scalebox{0.9}{
		\begin{tabular}{r l}
			\toprule
			{\bf Notation} & {\bf Description}\\
			\hline
			$N$ & Size of the population in the raw data\\
			$D$ & NDV of the raw data\\
			$q$ & Sample rate\\
			$n$ & Size of the population in the sample\\
			$k$ & Number of machines\\
			$n^{(1)},\ldots,n^{(k)}$ & Number of items in the sample of machine $k$\\
			$n_j$ & Element $j$ appear $n_j$ times in the sample\\
			$N_j$ & Element $j$ appear $N_j$ times in the population\\
			$\bar{N}$ &Average class size in population, $\frac{N}{D}$\\
			$d$ & NDV of total samples\\
			$d^{(1)},\ldots,d^{(k)}$ & NDV of sample in machine $k$ \\
			$F_1,\ldots,F_i$ & Number of items appear $i$ times in total population\\
			$f_0,f_1,\ldots,f_i$ & Number of items appear $i$ times in all sample\\
			$f_i(X)$ & Return the elements appear $i$ times in vector $X$\\
			$X_{=i}$ & The elements whose value is equal to $i$ in the stream $X$\\
			$X_{\neq i}$ & The elements whose value is not equal to $i$ in the stream $X$\\
			$\#\{\cdot\}$ & Size of set\\
			$s$ & Skewness parameter for a Zipfian population\\
			$\lambda$ & Mean of Poisson distribution\\
			$b$ & The parameter of HyperLogLog\\
			\bottomrule
		\end{tabular}
		}
		\vspace{-1em}
		\label{tab:notation}
	\end{small}
\end{table}
\subsection{Sampling-based NDV Estimators}

The NDV estimation problem aims to estimate the number of distinct values $D$ for a given data set with $N$ elements. 
Sampling-based NDV estimators~\cite{haas1998estimating,charikar2000towards,motwani2006distinct} take a small random sample of size $n$ from the data and approximate $D$ with pre-defined estimators. Under various assumptions on the data distribution, these methods provide empirical or theoretical guarantees for the quality of the estimators. For example, Motwani et al.~\cite{motwani2006distinct} propose an estimator that works well on data that follows the Zipfian distribution. In general, these estimators take the form of a function of the {\em frequency of frequency}, which is defined as follows. 


\header{\bf {\em Frequency of frequency}. }
The {\em frequency of frequency} of the raw data is composed of $F_i$, where $F_i$ is the number of elements that appear $i$ times in the raw data.


\header{\bf The number of unique values. }
We also define a sample's {\em frequency of frequency} as the set of $f_i$'s, where $f_i$ is the number of elements that appear $i$ times in the sample. For convenience, $f_1$ is frequently used, naming it the number of unique values.

\header{\bf Estimators from the Literature. }
Given the {\em frequency of frequency} of samples, one can estimate the NDV of the raw data with various sampling-based NDV estimators. We review some of the estimators that will be extended in our distributed framework, including GEE~\cite{charikar2000towards}, Shlosser~\cite{shlosser1981estimation}, Chao~\cite{ozsoyoglu1991estimating}, CL1~\cite{chao1992estimating}. These estimators were proposed in the database and statistical literature. We also modify some of them to better fit our framework in the following section. The formulas of these estimators can also be found in Table~\ref{tab:esti}. 
\begin{itemize}[leftmargin = *]
\item  {\em Guaranteed Error Estimator. }Charikar et al.~\cite{charikar2000towards} give a hard case for estimating the distinct value in a table column.~\cite{charikar2000towards} analyze the optimal error for the hard case and obtain the estimator as follows,
\begin{equation}\label{eq:gee}
\hat{D}_{GEE} = \sqrt{N/n} f_1 + \sum_{i=2} f_i. 
\end{equation}
In order to facilitate understanding and calculation, the GEE estimator can be rewritten by $d$ and $f_1$:
\begin{equation}\label{eq:gee2}
\hat{D}_{GEE} =d+\left(\sqrt{N/n}-1\right)f_1,
\end{equation}
where $d$ is the NDV of the sample.
\item {\em Chao's Estimator. }According to~\cite{chao1984nonparametric}, Ozsoyoglu et al.~\cite{ozsoyoglu1991estimating} apply the estimator $\hat{D}_{Chao}=d+\frac{f_1^2}{2f_2}$ to database NDV estimation. While $f_2 = 0$, $\hat{D}_{Chao}$ will blow up. 
With the Assumption\ref{ass-1} in the analysis part, the majority of elements appear once in the sample. Except for $f_1$, $f_2$ accounts for the vast majority of distinct values. It is reasonable for us to replace $f_2$ with $d-f_1$. Then we have a new estimator of Chao as:
\begin{equation}\label{eq:chao3}
\hat{D}_{Chao3}=d+\frac{1}{2}f_1^2/(d-f_1).     
\end{equation}
\item {\em Chao Lee's Estimator. }Chao Lee's Estimators are the extension of Chao's estimator. We take the simplest one in Chao Lee's estimation family. The complete derivation process can be found in Appendix~\ref{A-Esti}. Chao Lee's first estimator can be written as:
\begin{equation}\label{eq:CL1}
\hat{D}_{CL1}=\frac{d+f_1\cdot \max\left\{ \frac{d\sum_i i(i-1)f_i}{ (1-f_1/n)(n^2-n-1) } ,0 \right\} }{1-f_1/n}. 
\end{equation}

\item {\em Shlosser's Estimator. }Shlosser~\cite{shlosser1981estimation} derives the estimator,
\begin{equation}\label{eq:sh}
\hat{D}_{Sh} = d + \frac{f_1\sum_i (1-q)^i f_i }{ \sum_i iq(1-q)^{i-1} f_i }. 
\end{equation}
This estimator was constructed for language dictionaries. It assumes that the population is large, the sampling fraction is non-negligible, and the proportions of classes in the sample reflect the population, namely $\frac{E[f_i]}{E[f_1]}\approx \frac{F_i}{F_1}$.
\end{itemize}


%
%
%

\subsection{Sampling-based NDV Estimation in Communication Complexity Model}
%
This subsection will give the formal definition of the distributed sampling-based NDV estimation problem. We assume that amounts of data are dispersed over many devices. The communication cost becomes the most crucial complexity parameter in such a complicated processing system.
To model the distributed environment, we adapt the {\em communication complexity model}~\cite{yao1979some,kushilevitz1997communication}, which solely focus the communication cost. We first define the frequency vector, which is used to tally the number of times objects appear.


\begin{definition}[Frequency vector]
The associated frequency vector for a sample $S=(s_1,s_2,\ldots,s_n)$ is $X=( x_{s_1},x_{s_2},\ldots,x_{s_d} )$, where $x_{s_i}$ denotes the frequency of element $s_i$. To simplify, we denote $X=( x_{s_1},x_{s_2},\ldots,x_{s_d})$ as $X=(x_1,\ldots,x_d)$. $f^X$ denotes the {\em frequency of frequency} of $X$.
\end{definition}
Let $\ell_p$ denote the $\ell_p$ norm of the frequency vector (for the sample), and we have the following correlation between vector norms and frequency of frequency. 
\begin{equation}\label{eq:fan}
    \|X\|_p^p = \sum_{i=1} i^p f^{X}_i.
\end{equation}

Combining data from multiple machines can be regarded as merging the frequency vectors across these machines. We can only transmit the frequency dictionaries of samples from multiple machines.
\begin{definition}[Frequency dictionaries]
The associated frequency dictionary for a frequency vector $X=( x_{s_1},x_{s_2},\ldots,x_{s_d} )$ is $FD_{X}=\{s_1:x_{s_1},s_2:x_{s_2},\ldots,s_d:x_{s_d}\mid x_{s_i}>0\}$.
\end{definition}

The communication complexity model provides a computational model that characterizes the communication cost of such operations. In particular, we first present a simplified two-party communication model that will serve as a particular case of the communication complexity model. 


\begin{definition}[Two-party communication complexity model]
	Alice has a frequency vector $X=( x_{1},x_{2},\ldots,x_{m} )$ of her sample, and Bob has a frequency vector $Y=( y_{1},y_{2},\ldots,y_{m})$ of his sample. The goal is to compute $f=(f_1,f_2,\ldots)$, the frequency of frequency for the union of Alice and Bob's samples, with the lowest possible communication between Alice and Bob. 
\end{definition}


In general, we assume that there are $k$ machines. A broader definition of our problem is given as follows.

\begin{figure}[t]
\centering
\includegraphics[width=1.0\columnwidth]{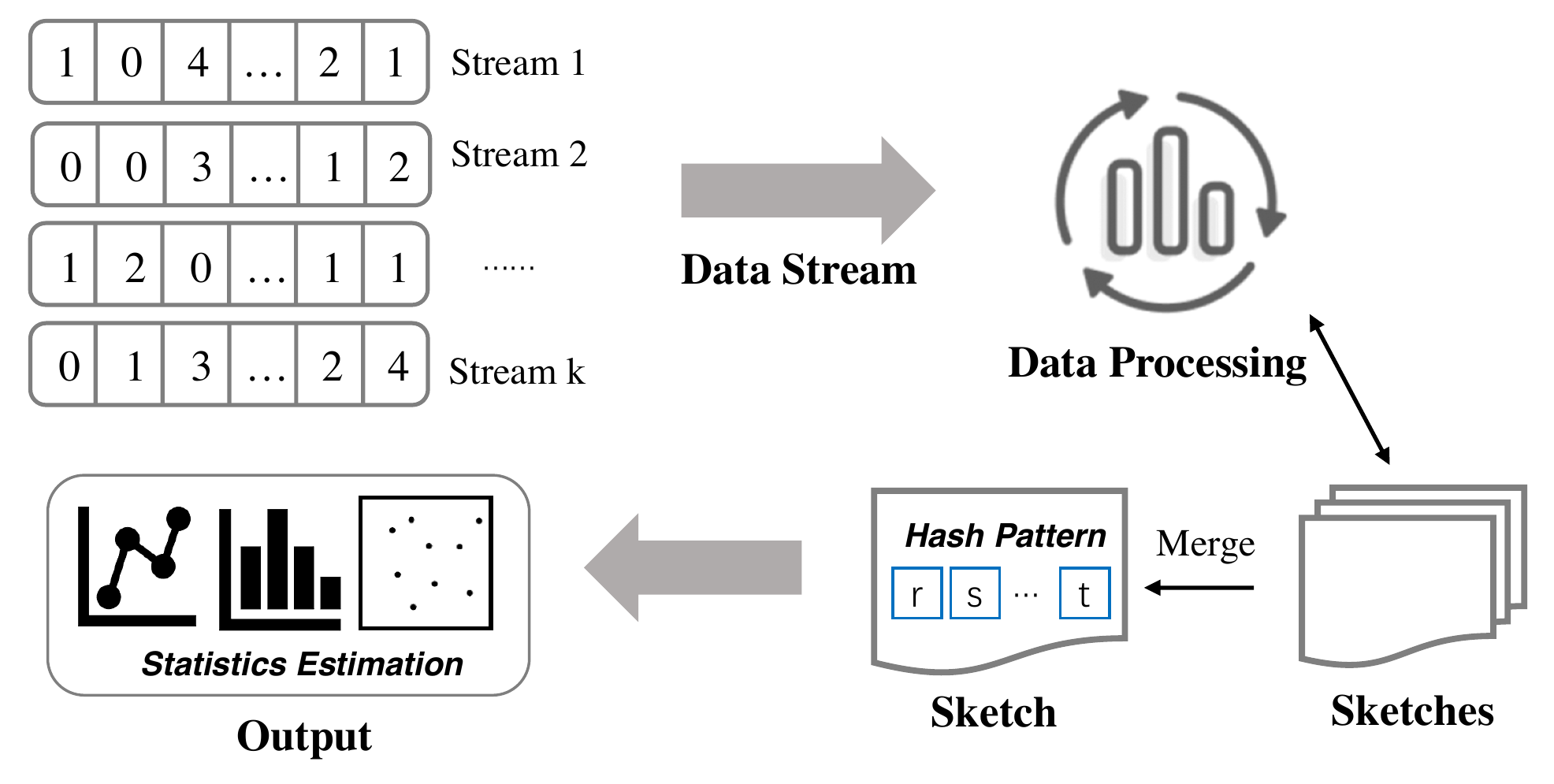}
\caption{Estimating with Streaming in Distributed Environment}
\label{fig:model2}
\vspace{-2em}
\end{figure}

\begin{definition}[Multi-party communication complexity model]
	Let $M_1,M_2,\ldots,M_k$ be a series of machines. Each machine has a frequency vector $X_{M}=(x_1,x_2,\ldots,x_m)$ of its sample. The goal is to return the {\em frequency of frequency} of the sample union in $k$ machines with the lowest possible communication cost. 
\end{definition}
After we obtain the frequency of frequency of the sample union, we can employ NDV estimators such as Equation~\eqref{eq:CL1} to approximate the NDV of the raw data union. 
Figure~\ref{fig:model2} shows how to use the communication complexity model to obtain NDV estimators in real-world applications. 
Note that after accepting the input sample from each machine, specific data actions such as hashing will be performed to extract sketches. Then, rather than receiving the complete data, the model will receive small sketches to save high transmission costs. Because sketches are mergeable, the mater node will combine all the sketches and deliver the estimation.



\subsection{Streaming algorithms}
Streaming models and sketching techniques are frequently used for processing massive amounts of data. Sketch techniques enable a low-memory single-pass estimation of the attributes of data items in a stream, such as heavy hitters~\cite{misra1982finding} and $\ell_p$ norms~\cite{indyk2006stable}. We now review some commonly utilized sketches in the streaming model.

\header{\bf $\ell_0$ norms estimation. }We employ an estimated distinct counting function int the data mining to return the NDV estimation. The HyperLogLog~\cite{flajolet2007hyperloglog} is used as the basic algorithm. HyperLogLog is a near-optimal technique for estimating distinct values. In order to deliver an exact answer, all the distinct values must be saved, which can consume a large amount of space. However, if we allow for a relative error to estimate the distinct values, we can still scan the data once and save a great deal of space in the process. Estimation's extremely compact data structure is referred to as sketches. There are numerous techniques for estimating unique values using sketches, including KMV~\cite{bar2002counting} and HyperLogLog~\cite{flajolet2007hyperloglog,heule2013hyperloglog}. These algorithms scan the data once, use a hashing mechanism and take up relatively little space. 
HyperLogLog is more accurate than KMV and will be used in the trials. 

\header{\bf $\ell_2$-norms estimation. }The estimator~\eqref{eq:CL1} has an item $\sum_i i^2f_i$ and computing all the $f_i$ while maintaining accuracy is difficult. Fortunately, $\sum_i i^2f_i$ is the $\ell_2$ norms of frequency vector $X$, and we have several algorithms to estimate $\ell_2$ norms in the streaming model. Except for $\ell_0$ estimation, $\ell_2$ estimation is the most extensively studied problem in $\ell_p$ estimation. Similar to $\ell_0$ estimation, $\ell_p$ estimation employs hashing to project the frequency and summarize the frequency moments using hash values, such as~\cite{coppersmith2004improved}. Indeed, these estimations of frequency moments concentrate on those elements that always exist, such as~\cite{alon1999space}. We primarily employ the $\ell_2$ norms in this research to estimate the distinct values of scaled data. The AMS Sketch~\cite{alon1999space} is the standard method for estimating the $\ell_2$ norms. Given parameters $\varepsilon$ and $\delta$, the summary employs space $\mathcal{O}(\frac{1}{\varepsilon^2} \log \frac{1}{\delta})$ and ensures that its estimate is within the relative $\varepsilon$ error of the actual $\ell_2$ norms with a probability of at least $1-\delta$. The Count Sketch~\cite{charikar2002finding} is thought of as an extension of the earlier AMS Sketch~\cite{alon1999space}. We use Count Sketch to calculate $\sum_i i^2f_i$ in experiments.

\section{Distributed sampling-based NDV estimation}
In this section, we concentrate on the sampling-based estimation of NDV in a distributed environment. We first present the negative results, which inspire us to consider additive error and introduce the mild assumption in Section 4.

\subsection{Negative Result}
\header{\bf Exact computation of $f_1$'s} Almost all estimators based on sampling requires $f_1$ to estimate NDV. For instance, GEE estimator requires simply $f_1$ and $d$. Therefore, the natural idea is to compute $f_1$'s of the sample union by delivering the dictionaries of frequency vectors across the machines. However, as data volume increases, transmitting the dictionary of frequency vectors incurs a high communication cost. For the data with many different values, each machine will have a massive frequency dictionary. We will incur high costs in sending the dictionaries. 



\header{\bf Approximating $f_1$'s with relative error. } In a distributed context, we can estimate $d$ with a relative error using the notion of $\ell_0$ sketch, such as HyperLogLog. The operations of HyperLogLog are like the operations of the set, except for intersections. Set intersection returns the common elements of two sets. Computing $f_1$ is equivalent to computing the size of the intersection of two sets. Calculating the $f_1$ of the two devices is the simplest case of merging two frequency dictionaries. We prove and establish Theorem~\ref{thm:1} that estimating the size of the set intersections can be reduced to computing the merging frequency dictionary of two machines.

\begin{theorem}\label{thm:1}
	Calculating the size of set intersection can be reduced to calculating the merging frequency dictionaries of two machines. Assume $m$ is the size of $A$ and $B$'s frequency vectors $Z$. Estimating $f_1(Z)$ with relative error $\varepsilon$, the communication complexity of estimation is $\Omega(m)$.
\end{theorem}

In other words, if we do not transmit the complete dictionary of samples, we cannot guarantee the relative error $f_1$ in all situations.

\subsection{Estimate $f_1$ with Small Communication}


We present an algorithm for reducing the communication cost of estimating $f_1$'s under mild assumptions. Before diving into technical details, we first present the high-level ideas of our method via a simple two machines distributed model.

\header{\bf Estimate $f_1$ on two machines. }
We now consider a two-party communication model where Alice and Bob each possess a sample with frequency vectors $X=( x_1,\ldots,x_m)$, and $Y=( y_1,\ldots,y_m)$, respectively. The goal is to estimate $f_1$ for the union $X+Y$ without transferring the entire sample $X+Y$. By sorting, Alice and Bob can obtain a frequency dictionary. According to Alice's dictionary, Alice can acquire vector $X_{=1}$ which contains the elements that appear just once, and the vector $X_{\neq 0}$, which has all of Alice's elements. The number of distinct items in the distributed environment will be maintained using the $\ell_0$ sketch, such as HyperLogLog. The precise count of distinct elements in vector $X$ can be expressed as $\| X \|_0$. Because the $\ell_0$ sketch supports the merge operation, $\|X+Y\|_0$ represents the number of distinct elements in the merge of Alice's and Bob's $\ell_0$ sketch. To calculate $f_1$, we make use of the following expression:
\begin{equation}\label{eq:f1}
f_1(X+Y) = \|X_{=1}+Y\|_0-\|Y\|_0 + \|X+Y_{=1}\|_0 - \|X\|_0.
\end{equation}
For proof, note that Equation~\eqref{eq:f1} is divided into two parts: $\|X_{=1}+Y\|_0-\|Y\|_0$ and $\|X+Y_{=1}\|_0 - \|X\|_0$. If we exchange $x$ and $y$, the two parts just swap values, and the formula remains the same. Thus, the analysis for the two parts is similar. We analyze the former first. With the definition of $\|\cdot\|_0$, we have \begin{equation}\label{eq:bridge}
\|X_{=1}+Y\|_0-\|Y\|_0=\#\{ X_{=1} \lor Y_{\neq 0}\} -\#\{Y_{\neq 0}\}. 
\end{equation}
With the set operations, we can derive
\begin{equation}\label{eq:set}
\#\{ X_{=1} \lor Y_{\neq 0}\} -\#\{Y_{\neq 0}\}=\#\{ (X_{=1} \lor Y_{\neq 0})\land Y_{=0}\}= \#\{ X_{=1} \land Y_{=0}\}.
\end{equation} 
$\#\{ X_{=1} \land Y_{=0}\}$ denotes the number of elements that appear once in $X$ but not in $Y$. So $f_1$ is equal to the sum of $\#\{ X_{=1} \land Y_{=0}\}$ and $\#\{ Y_{=1} \land X_{=0}\}$.
Therefore, combining Equation~\eqref{eq:bridge}~\eqref{eq:set}, we can derive
$$
\|X_{=1}+Y\|_0-\|Y\|_0 + \|X+Y_{=1}\|_0 - \|X\|_0=f_1(X+Y).
$$
The calculated expression of $f_1$ is not unique. However, the expression above shows the connection between the set operation and the calculation of $f_1$. Next, we extend the process of calculating $f_1$ from two machines to multi-machines.

\begin{algorithm}[t]
	\caption{ Construction of Pre-Merged Sketches  \label{alg-Pre} }
	\KwIn{ NDVSketch: Array of the sketches to store the distinct values of k machines.}
	\KwOut{PMSketch: The Pre-Merged sketches array to store the intermediate merged sketches from different machines.}
	PMSketch[$0$]$=$NDVSketch \;
	$n$ $\leftarrow$ NDVSketch.length\;
	$l$ $\leftarrow$ 0\;
	\While{$n>2$}{
	    $n \leftarrow n$/2\;
	    \For{$i$ from 0 to $n$}{
	        Initialize $\ell_0$ sketch tempS with PMSketch[$l$][$2*i$] \;
	        tempS.merge(PMSketch[$l$][$2*i+1$])\;
	        PMSketch[$l+1$].append(tempS)\;
	    }
	}
	\Return\;
\end{algorithm}

\header{\bf Extension: Estimate $f_1$ on Multi-Machines. }Consider the extension of estimating $f_1$ about two machines. If we have more than two machines, we can still use the combination of $\ell_0$ norms to calculate the $f_1$ of these machines. $\mathcal{M}=[M_1,M_2,\ldots,M_k]$ is a machine list. $\ell_0$ sketches are similar to sketches of sets that only support union operations in our framework. Considering arbitrary machine $M_j$, we just need to count the number of items that only appear in this machine. Then the sum of all items that only appear in one machine is $f_1$. 

For an arbitrary machine $X$, we summarize the frequency vectors of all the machines except $X$ as $Y$. Then we can use the Equation~\eqref{eq:bridge} to count the number of items that only appear in machine $X$. Equation~\eqref{eq:bridge}'s left part represents the number of items that only appear in machine $X$. In the right part, $\|X_{=1}+Y\|_0$ represents the number of items that appear once in $X$ or other machines, and $\|Y\|_0$ represents the number of items in other machines. We are both using the $\ell_0$ sketch for estimation.

However, if we traverse all the machine cases and merge the other machines for each machine, the time complexity is $O(k^2)$, limiting the system's scalability. Fortunately, many merge operations are duplicated. We use a binary tree to store the intermediate merged sketches, constructed with $O(k)$ space and $O(k)$ time complexity. Algorithm~\ref{alg-Pre} provides the algorithm and pseudocode for constructing the intermediate merged sketches from different machines.

According to the Pre-Merged Sketches, we give the algorithm~\ref{alg-Estif1} and pseudocode to estimate $f_1$. The level of sketches is $O(\log k)$ and we need at most 2 sketches per level to cover all machines.

\begin{algorithm}[t]
	\caption{ $Esti F_1$  \label{alg-Estif1} }
	\KwIn{ PMSketch: The Pre-Merged sketches array to store the intermediate merged sketches from different machines.
	    F1Sketch: Array of the sketches to store the unique values of $k$ machines, namely $f_1$ of each machine.
	    $k$: The number of machines}
	\KwOut{The estimation of $f_1$}
	\SetKw{XOR}{XOR}
    Initialize $f_1$ $\leftarrow$ $0$\;
	\For{$index$ from $0$ to $k$}{
	   Initialize $B$ with empty $\ell_0$ Sketch \; 
	  \For{$l$ from $0$ to PMSketch.height-1}{
	  // From down to top, merge the PMSketch\;
	        \If{$l=0$}{
	            $B$ Merge PMSketch[$l$][$index$ \XOR 1] \;
	            $Next$ $\leftarrow$ (($index$ \XOR 1)/2) \XOR 1  \;
	        }\Else{
	            $B$ Merge PMSketch[$l$][$Next$] \;
	            $Next$ $\leftarrow$ ($Next$/2) \XOR 1 \;
	        } 
	  }
	  $B$ merges the rest PMSketch's sketches which have not merged PMSketch[0][$index$] and have not been merged by other PMSketch's sketches\;
	  $f_1$ $\leftarrow$ $f_1$-$B$.estimate() \;
	  $B$ merges F1Sketch[$index$]\;
	  $f_1$ $\leftarrow$ $f_1$+$B$.estimate() \;
}
	\Return $f_1$ \;
\end{algorithm}

\subsection{Coupling with Existing NDV Estimation}
In this section, we present how our framework can be coupled with the different traditional estimators. When the order of $f_i$ is high, the computational complexity cannot be linear. However, we can avoid calculating the high-frequency term by modifying the estimators.



\header{\bf Estimate GEE and Chao with $f_1$ and $d$. }For some basic estimators, according to Equation~\eqref{eq:gee2} and Equation~\eqref{eq:chao3}, it is enough to estimate NDV with $f_1$ and $d$. 

\header{\bf Estimate Chao Lee's Estimator with $\ell_2$ Norms. }For some complicated estimators, they consider the influence of all $f_i$. Estimating all of $f_i$ with high precision is difficult. However, we can estimate the $\ell_p$ norms of frequency vector with some stream models. We observe that $\sum i^p f_i$ can be summarized as $\ell_p$ norms of $f$ for these complex estimators. We use the $\hat{D}_{CL1}$ as an example to show how to adjust the estimators. The relationship between $\|X\|_p^p$ and $\sum_i i^p f_i$ is given by Equation~\eqref{eq:fan}. Targeted at Equation~\eqref{eq:CL1}, we have a new expression for $\hat{D}_{CL1}$
\begin{equation}\label{eq:CL1ad}
\hat{D}_{CL1-Adjust}=\frac{d+f_1\cdot \max\left\{ \frac{d\cdot(\|f\|_2^2-n) }{ (1-f_1/n)(n^2-n-1) } ,0 \right\} }{1-f_1/n}.
\end{equation}
We use Count Sketch with a relative error to estimate $\|X\|_2^2$. $\|X\|_2^2$ is sufficient to deal with the majority of complicate estimators.

\header{\bf Estimate Shlosser's Estimator with resampling. }We can deduce the meaning of estimators and propose the equivalent forms of estimators that cannot be transformed into $\ell_p$ norms. Shlosser's estimator assumes that the population size is large and $\frac{E[f_i]}{E[f_1]}\approx \frac{F_i}{F_1}.$ Then we have $F_0\approx\frac{F_1\cdot E[f_0]}{E[f_1]}$. When the population size is large, the sampling procedure can be seen as sampling from a Binomial distribution at a fixed sample rate $q$. Then we have $E[f_0]=\sum_i (1-q)^if_i$ and $E[f_1]=\sum_iiq(1-q)^{i-1}f_i$.  There is no doubt that it is difficult to calculate each $f_i$ and then return Shlosser's estimator. Nevertheless, we just need the expectation of $f_0$ and $f_1$, which can be computed by resampling. So we have the adjusted estimator of Shlosser as follows.
$$\hat{D}_{Sh-Adjust}=d+\frac{f_1 \cdot E[f_0]}{E[f_1]}=d+\frac{f_1\cdot (d-d^{resample})}{f_1^{resample}}.$$
We summarize the original and adjusted forms of NDV estimators in Table~\ref{tab:esti}. According to Table~\ref{tab:esti}, we only require $f_1$, $d$ and $\|f\|_2^2$ for calculating sampling-based NDV.
\begin{table}[t]
		\centering
		\caption{Estimators and their Approximation.}
		\vspace{-1em}
		\scalebox{0.9}{
		\begin{tabular}{r c c}
			\toprule
			\hline
			Estimator & Original Expression & Adjusted Expression \\
			\hline
			$\hat{D}_{GEE}$ & $ \sqrt{\frac{N}{n}} f_1 + \sum_{i=2} f_i $&$\hat{d}+\left(\sqrt{\frac{N}{n}-1}\right)\hat{f}_1$  \\
			$\hat{D}_{Chao2}$ & $d+\frac{f_1(f_1-1)}{2(f_2+1)}$&$\hat{d}+\frac{\hat{f}_1^2}{2(\hat{d}-\hat{f}_1)}$\\ 
			$\hat{D}_{CL1}$ & $\frac{d+f_1\cdot \max\left\{ \frac{d\sum_i i(i-1)f_i}{ (1-f_1/n)(n^2-n-1) } ,0 \right\} }{1-f_1/n}$&$\frac{\hat{d}+\hat{f}_1\cdot \max\left\{ \frac{\hat{d}\cdot(\|X\|_2^2-n) }{ (1-\hat{f}_1/n)(n^2-n-1) } ,0 \right\} }{1-\hat{f}_1/n}$ \\
			$\hat{D}_{Sh}$ &$\hat{d} + \frac{f_1\sum_i (1-q)^i f_i }{ \sum_i iq(1-q)^{i-1} f_i }$&$d+\frac{\hat{f}_1\cdot (\hat{d}-d^{resample})}{f_1^{resample}}$ \\
			\bottomrule
		\end{tabular}
		}
		\vspace{-1em}
		\label{tab:esti}
\end{table}

\section{Analysis}
In this section, we first give a mild assumption about the data distribution. Next, we will provide a theoretical demonstration that we will have a relative error under this assumption. Besides, we analyze the communication cost of our algorithm to evaluate the GEE estimator. We will also give a theoretical analysis of the error in GEE. 

\header{\bf Assumption on Distribution. }If we want to count $f_1$ of samples accurately, the communication cost is determined by the distinct values of samples. Without any prior hypothesis, we suppose that data is distributed evenly among each machine. Our estimators focus on estimating $d$ and $f_1$, according to Table~\ref{tab:esti}. The following is the mild assumption.

\newtheorem{assumption}{Assumption}[section]
\begin{assumption}\label{ass-1}
If we uniformly sample data from some known distributions or real-world datasets with a small sample ratio $q$, we have $f_1 \geq c \cdot d, c\in (0,1)$, where $c$ will not be too small, implying that the majority of elements in the sample will appear only once.
\end{assumption}

Because some distribution's $f_1/d$ is close to zero, implying that $f_1$ and $d$ are small, we can directly transfer the frequency dictionaries. In fact, in any case, we should calculate the frequency dictionaries at first. So we can find out whether the sample meets the Assumption~\ref{ass-1} when obtaining the frequency dictionary on each machine.

Besides, under different distributions with fixed parameters, we can adjust the sample rates to achieve $f_1\geq c\cdot d$. To simplify the calculation, we use $Poi(iq)$ to approximate $Binomial(i,q)$, where $q$ is the sample ratio because we have a large population. Then, for a series of $F_i$, we have the following expression for $f_1$.
$$
E[f_1]=\sum_i iq e^{-iq}\cdot F_i.
$$
To compute the distinct values of the sample, we have the expression $E[d]=E[D-f_0]=\sum_i F_i(1-e^{iq})$. Combining the calculation expressions of $E[f_1]$, $E[d]$, and $f_1/d\geq c$, we should solve the inequation:
\begin{equation}\label{eq:error}
\frac{\sum iq\cdot e^{-iq}F_i }{\sum F_i (1-e^{-iq}) }\geq c.
\end{equation}
For any distribution, we can use Equation~\eqref{eq:error} to determine $q$ or $c$. Even when $c$ takes a relatively small value, we still can guarantee the accuracy. For example, as our experiment has shown, in realistic data, $f_1/d\approx 0.14$(LO\_ORDERKEY), 0.0028(LO\_ORDTOTALPRICE), and 0.0018(LO\_REVENUE) are also acceptable. We need to control the accuracy of HyperLogLog. After this preparation, we will analyze the error of estimating $f_1$ in the next part.

\header{\bf Error of $f_1$ Estimation.} 
All the HyperLogLog estimators will be bound by $\varepsilon_d$. One HyperLogLog requires $O(1/\varepsilon_d^2\log(1/\delta)\log\log n )$ bits with the median trick according to~\cite{cormode2020small}, where $\delta$ is the probability of exceeding the bound. The theorem is as follows.
\begin{theorem}\label{the-1}
The given data uniformly distributed in $k$ machines meets the Assumption~\ref{ass-1} with parameter $c$. Using the HyperLogLog($\delta,n$) with relative error $\epsilon_d$ as $\ell_0$ sketch, we only need to send two sketches for each machine. The total communication cost of Algorithm~\ref{alg-Estif1} is $$O(k/\varepsilon^2_d \log(1/\delta^2)\log\log(n),$$ which takes $O(k\log k)$ merge operations and $k$ HyperLogLog($\delta,n$).
\end{theorem}


\header{\bf Error of GEE Estimator. }Charikar et al.~\cite{charikar2000towards} provide a lower bound on the ratio error for sample-based NDV estimators. Formally, the {\em ratio error} of an estimation $\hat{D}$ w.r.t the genuine $D$ is
\begin{equation}
    Error(\hat{D},D)=\max\{\hat{D}/D,D/\hat{D}\}.
\end{equation}

Charikar et al.~\cite{charikar2000towards} also give and prove the fact as follows.
\newtheorem{fact}{Fact}[section]
\begin{fact}\label{the-2}
The expected ratio error of GEE is $O(\sqrt{n/r})$ when it samples $r$ values from any possible input of size $n$.
\end{fact}
According to Theorem~\ref{the-1} and Fact~\ref{the-2}, we also have the ratio error guarantee for $\hat{D}_{GEE}$ computed by our algorithms as follows.

\begin{theorem}\label{the-3}
The expected ratio error of GEE is $O(\varepsilon\sqrt{n/r})$ when it samples $r$ values from any possible input of size $n$ and is computed by Algorithm~\ref{alg-Estif1} with relative error $\varepsilon$.
\end{theorem}

\header{\bf Remark. }The reason we require Assumption\ref{ass-1} is that, without any assumption, there is no relative-error estimation for $f_1$. We have proved that the set intersection problem can be reduced to the problem of merging frequency dictionaries. We only need to show that there is no relative error estimation for the size of the set intersection.

For the size of set intersection, we have the invariant as follows,
\begin{equation}
\big| A\cap B\big|=\big|A\big| + \big| B\big| - \big| A\cup B \big|.  \label{eq-inter}
\end{equation}

According to Equation~\eqref{eq-inter}, if we return $\big| A\cup B \big|$ with a relative error $\varepsilon\big| A\cup B \big|$, the lower bound on $\big| A\cap B \big|$ estimation error is $\big|A\big| + \big| B\big|-\varepsilon \big| A\cup B \big|$. As a result, we will not know the relative inaccuracy of all possible scenarios. For instance, consider the case where we control the relative error rate $\varepsilon'$, and there is no intersection between $A$ and $B$. As long as $\varepsilon'$ is not equal to 0, for $\big| A\cup B \big|$, then the intersection of $A$ and $B$ will always have a relative infinity error. While set intersection does not support relative error estimation, it appears that we can still manage the absolute error in some range for set intersection problems and $f_1$ estimation. The key to minimizing the relative error of set intersections is ensuring that the size of the intersection is sufficiently large. For example, if the size of the set intersection $c\dot | A \cap B|$ equals $| A \cup B|$, the set intersection error will be $c\cdot | A \cup B|=c\varepsilon' |A\cap B|$. The same holds for the assessment of $f_1$.

\section{Experiment}
In this section, we experimentally evaluate our method in different distributed environments. Our code is publicly available.\footnote{\url{https://github.com/llijiajun/NDV_Estimation_in_distributed_environment.git}}

\header{\bf Datasets. } We evaluate our experiments on  two synthetic datasets and one real-world dataset, the star schema benchmark(SSB)~\cite{o2009star}. One of the synthetic datasets follows the Poisson distribution with different mean (e.g., [50, 100, 200]), and the other follows the Zipfian distribution with different skewness factors (e.g., [1.2, 1.5 and 2]). Additionally, We modify the population size of synthetic data to manipulate the data's scale. The size of the population ranges from $10^{9}$ to $10^{12}$. We employ three columns of the star schema benchmark(SSB) dataset: LO\_ORDERKEY, LO\_ORDTOTALPRICE, and LO\_REVENUE, abbreviations for orderkey, totalprice and revenue. We use the fact table with a scaling factor of 15,000, resulting in about 900M rows of sample data. Figure~\ref{fig:all}(a), (c), and (e) depict the $F_i$'s values of three columns. we do not generate or store the raw data to simulate a large-scale dataset. Instead, we produce $F_i$'s, the {\em frequency of frequency} of the raw data, and sample from them. 
To imitate distributed environments on a single workstation, we divide the sample data into 1,024 files as the different workers. The partition of data determines the worker on Spark.

\subsection{Simulated Experiment}
We begin by simulating the experiments on a single machine to evaluate the communication cost incurred by various distributed NDV methods. The experiment is performed on a Linux machine with an Intel Xeon CPU clocked at 2.70GHz and 500 GB memory. All sketch algorithms are implemented in C++, including HyperLogLog and Count Sketch. 

\header{\bf Methods. } We compare our Algorithms~\ref{alg-Estif1}, $Estif_1$ with $Exactf_i$, which corresponds to directly merging the frequency dictionaries to obtain the precise $f_i$'s. We use the $\hat{f}_1$ and $f_i$'s calculate by two methods to form four classic NDV estimators, including Chao's~\cite{ozsoyoglu1991estimating}, Shlosser's~\cite{shlosser1981estimation}, CL1's~\cite{chao1992estimating} estimator, and GEE~\cite{charikar2000towards}. 


\begin{figure}[t]
	\centering
	\vspace{-1em}
	\includegraphics[width=80mm]{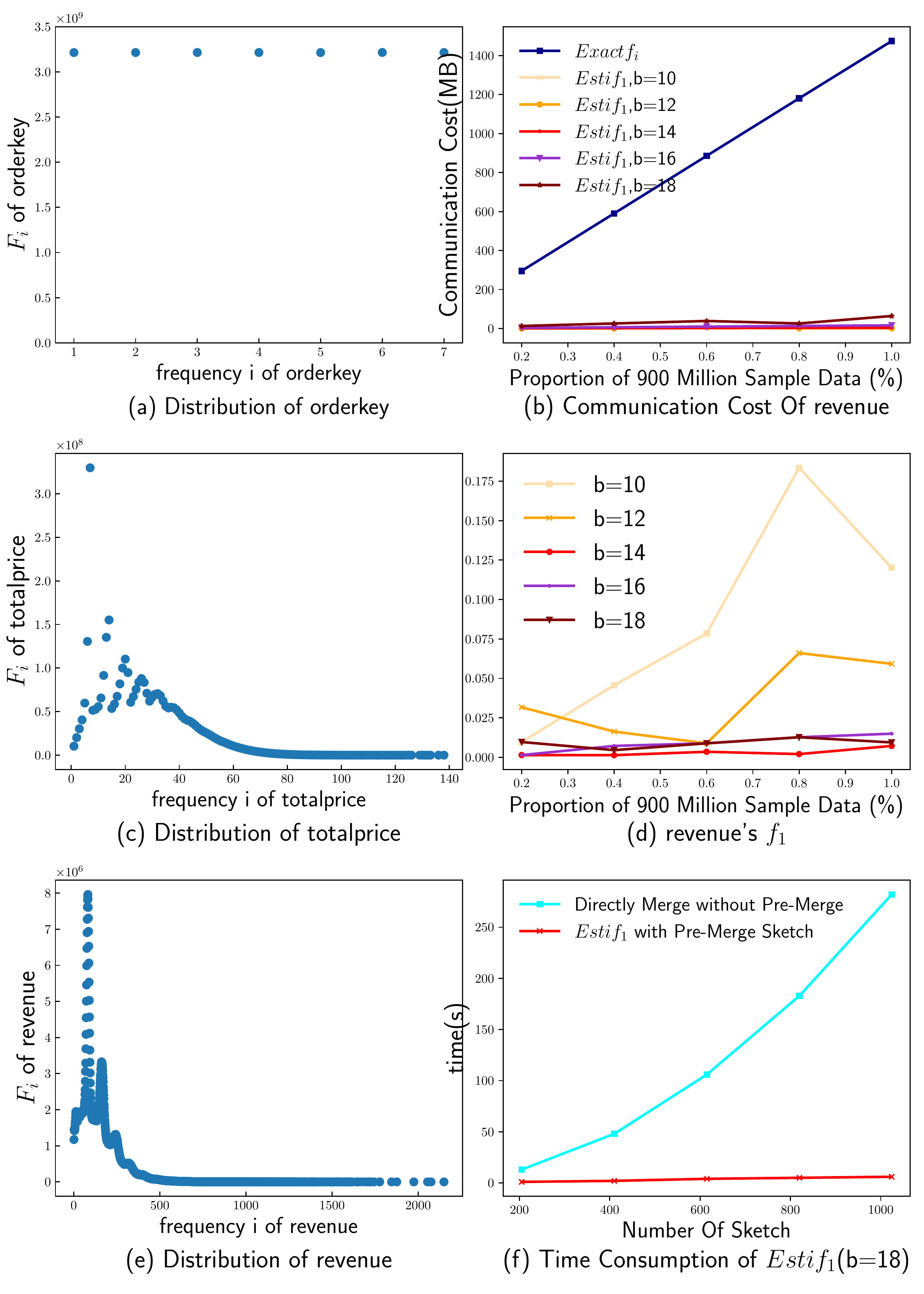}
	\vspace{-1em}
	\caption{Distribution of Real-World datasets and Communication Cost of Different Methods with Different Parameter. b is the parameter of HyperLogLog.}
	\label{fig:all}
	\vspace{-1em}
\end{figure}



\header{\bf Parameter Setup.} We evaluate all sampling methods with a sample rate of 0.01. Our experiments explore a variety of parameter $b$ (e.g. [10, 12, 14, 16, 18]) of HyperLogLog's. We use $\gamma=0.1$ and $\epsilon_{CS}=0.01$ to implement the Count Sketch. 

\begin{figure}[t]
	\centering
    \hspace{-4mm}\includegraphics[width=45mm]{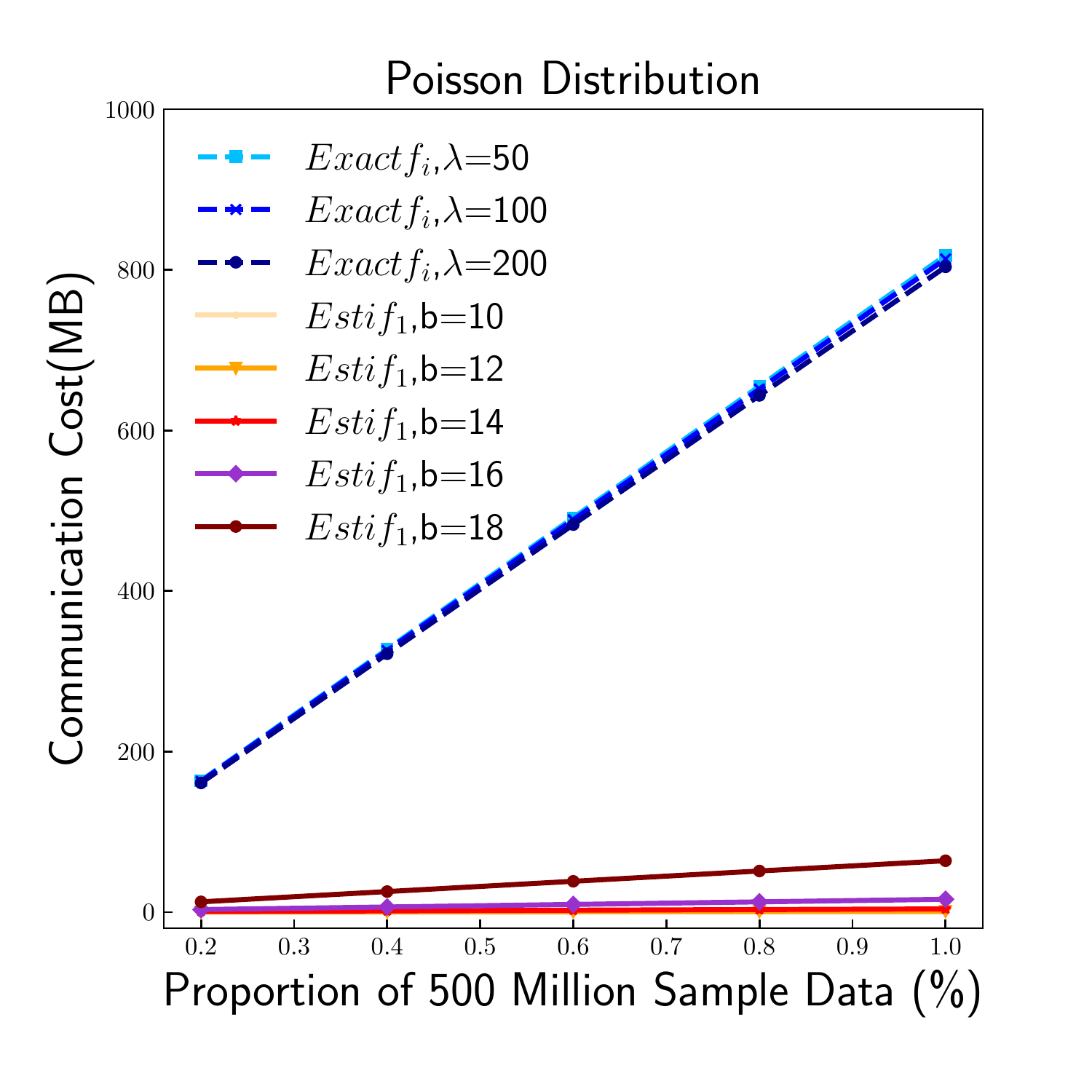}
    \hspace{-6mm}\includegraphics[width=45mm]{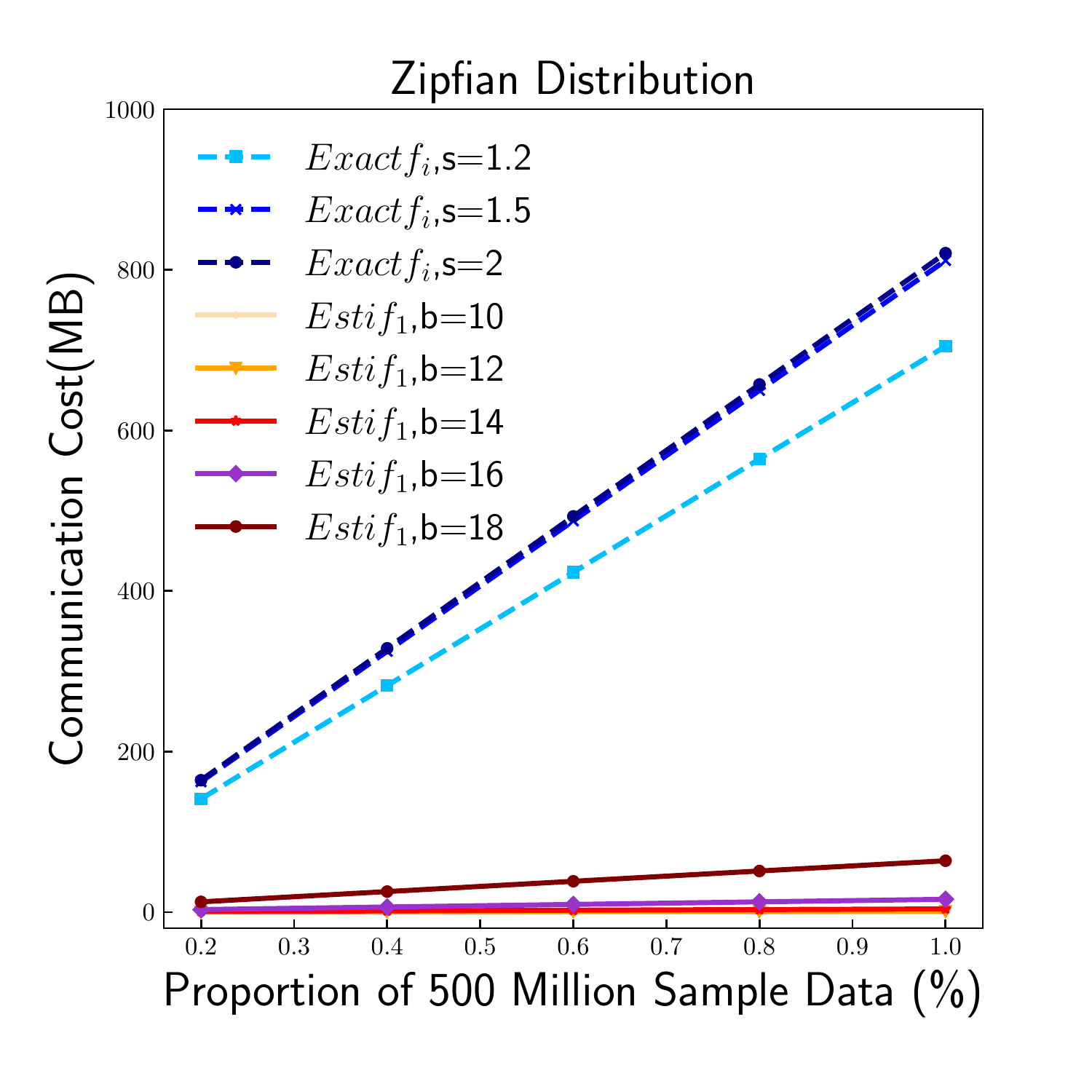}
    	\vspace{-1em}
	\caption{Communication Cost of Different Methods with Different Parameter. $\lambda$ is the parameter of Poisson distribution; s is the parameter of Zipfian distribution; $b$ is the parameter of HyperLogLog.}
		\vspace{-1em}
	\label{fig:communication}
\end{figure}

\header{\bf Results. }We progressively merge files on a single machine to simulate a multi-party communication complexity model. We make the following observations:
\begin{itemize}[leftmargin = *]
\vspace{0 mm}
\item {\em Communication. }We record simulated communication costs, which are the size of sketches or dictionaries transferred from a workers. Figure~\ref{fig:communication} and Figure~\ref{fig:all}(b) show the simulated communication cost obtained by $Exact f_i$ and our method $Esti f_1$. 
The communication cost of our method is unaffected by data size or distribution parameters, whether synthetic or real-world data. The communication cost of our method is only determined by the sketch's parameters and the number of machines. 
We show the communication cost of revenue in Figure~\ref{fig:all} because revenue's $f_1/d$ is the smallest of three real-world data. The rest results will be shown in Appendix~\ref{app-exp}. 
\item {\em Performance. }The performance of $Esti f_1$ and the adjusted estimators vary on different datasets. Figure~\ref{fig:all}(d) shows that using HyperLogLog with parameter $b\leq 12$, which corresponds to an error ratio, 0.016, we can estimate $f_1$ with a relative error less than 0.1. The values without parentheses in Table~\ref{tab:exp1.1} are the relative error between our estimators implemented by $Esit f_1$ and the original estimators implemented by $Exact f_i$. With the increase of b, we have a more accurate estimation of the original estimators, but we do not necessarily get closer to the actual values of $D$. Because some estimators, such as $\hat{D}_{Shlosser}$ for Poisson distribution, do not perform well, our adjusted estimators also do not perform well. In the case of our experiments, $\hat{D}_{GEE}$ and $\hat{D}_{CL1}$ are more stable and accurate for NDV estimation. In real applications, a more appropriate estimator can eventually be used, such as $D_{Chao}$. As long as the relatively better estimator uses $f_1$, we can approximate $D$ with a tolerable error.
\item {\em Scalability. }To make sure our algorithm is scalable, we count the time of $Esti f_1$ with the growth of machines. When we have thousands of machines and HyperLoglog's parameter $b=18$, the time of $Esti f_1$ will not be less than 10 seconds, as shown in Figure~\ref{fig:all}(f). Figure~\ref{fig:all}(f) also shows the consuming time of $Esti f_1$ with and without Pre-Merge. The blue line in Figure~\ref{fig:all}(f) denotes the time of $Esti f_i$ without Pre-Merge which complexity is $O(k^2)$. The red line in Figure~\ref{fig:all}(f) denotes the time of $Esti f_i$ with Pre-Merge, which argues that Algorithm~\ref{alg-Pre} is necessary.
\end{itemize}

\begin{table}[t]
	\begin{small}
		\centering
		\caption{Relative Error of Our Methods with different parameters on 500 Million Sample Data. Values without parentheses denote the relative error about estimators and values with parentheses denote the relative error about true $D$ }
			\vspace{-1em}
		\scalebox{0.9}{
		\begin{tabular}{l c c c c  }
			\toprule
			\hline
		Parameter  & $\hat{D}_{GEE}(\hat{D})$ & $\hat{D}_{Chao2}(\hat{D})$&$\hat{D}_{Sh}(\hat{D})$&$\hat{D}_{CL1}(\hat{D})$\\
			\hline
		b=10, $\lambda=50$&0.072(1.9)&0.24(0.24)&0.12(20)&0.13(0.54) \\
		b=14, $\lambda=50$&0.01(2.1)&0.13(0.13)&0.027(22)&0.089(0.93)\\
		b=18, $\lambda=50$&0.0092(2.1)&0.12(0.12)&0.028(22)&0.059(0.88) \\
		b=10, $\lambda=100$&0.034(2.8)&0.14(0.14)&0.077(21)&0.019(0.56)\\
		b=14, $\lambda=100$ &0.0057(2.9)&0.12(0.12)&0.033(22)&0.047(0.66)\\
		b=18, $\lambda=100$ &0.013(2.9)&0.12(0.12)&0.044(22)&0.018(0.61)\\
		b=10, $\lambda=200$& 0.083(2)&0.1(0.1)&0.18(9.3)&0.049(0.25)\\
		b=14, $\lambda=200$ & 0.0032(2.3)&0.078(0.078)&0.068(11)&0.11(0.17)\\
		b=18, $\lambda=200$ &0.0073(2.3)&0.077(0.077)&0.08(11)&0.011(0.33)\\
		b=10, $s=1.2$ &0.011(0.78)&0.34(0.87)&0.04(0.69)&0.032(0.048) \\
		b=14, $s=1.2$ &0.015(0.77)&0.27(0.86)&0.0034(0.77)&0.14(0.12)\\
		b=18, $s=1.2$ & 0.01(0.78)&0.4(0.88)&0.029(0.71)&0.041(0.057)\\
		b=10, $s=1.5$ &0.052(0.84)&0.6(0.88)&0.086(0.46)&0.44(0.5) \\
		b=14, $s=1.5$ &0.0092(0.83)&0.42(0.83)&0.02(0.56)&0.018(0.1)\\
		b=18, $s=1.5$ &0.0078(0.83)&0.37(0.82)&0.021(0.56)&0.3(0.38) \\
		b=10, $s=2$ &0.014(0.87)&0.47(0.041)&0.041(0.26)&0.38(0.97)\\
		b=14, $s=2$&0.0048(0.87)&0.26(0.1)&0.0029(0.31)&3.3(5.1)\\
		b=18, $s=2$ &0.0082(0.87)&0.47(0.62)&0.021(0.29)&0.62(1.3)\\
		b=10, orderkey&0.019(0.62)&0.4(0.4)&0.054(2.6)&0.76(2.5)\\
		b=14, orderkey&0.023(0.62)&0.43(0.43)&0.045(2.6)&0.26(1.5)\\
		b=18, orderkey&0.0026(0.62)&0.08(0.082)&0.014(2.7)&2.2(5.4)\\
		b=10, totalprice&0.092(0.98)&0.77(0.38)&0.1(16)&0.93(1.7)\\
		b=14, totalprice&0.011(0.8)&0.13(0.32)&0.03(14)&0.2(0.71)\\
		b=18, totalprice&0.0041(0.81)&0.1(0.3)&0.02(14)&0.25(0.78)\\
		b=10, revenue&0.099(1.9)&0.18(0.27)&0.18(10)&0.06(0.19)\\
		b=14, revenue&0.0039(2.2)&0.13(0.23)&0.042(12)&0.038(0.31)\\
		b=18, revenue&0.0085(2.2)&0.14(0.24)&0.059(12)&0.015(0.25)\\
			\bottomrule
		\end{tabular}}
		\label{tab:exp1.1}
		\end{small}
			\vspace{-1em}
\end{table}

\subsection{Experiments on Spark}

In this section, we deploy the sampling-based estimation of NDV in a real-world distributed environment and test the practical running time efficiency and I/O cost in the distributed environment. Because we use the same data, the performance on Spark is the same as in simulated experiments. We mainly compare the time cost and the I/O cost. 

\header{\bf Experimental setting.} The distributed environment includes sixteen machines with Intel(R) Xeon(R) CPUs clocked at 3.10GHz, 64 GB memory and 500 GB disk space. We implement the $Exact f_i$ algorithm, which calculates the {\em frequency of frequency} of sample with the MapReduce framework. We also implement our algorithm, which estimates the {\em frequency of frequency} of a sample with a sketch based on Map Reduce. The source codes on the distributed machines are implemented by PySpark and Cython on Spark version 3.1 and Hadoop version 3.3. 

\begin{figure}[t]
	\centering
	\includegraphics[width=86mm]{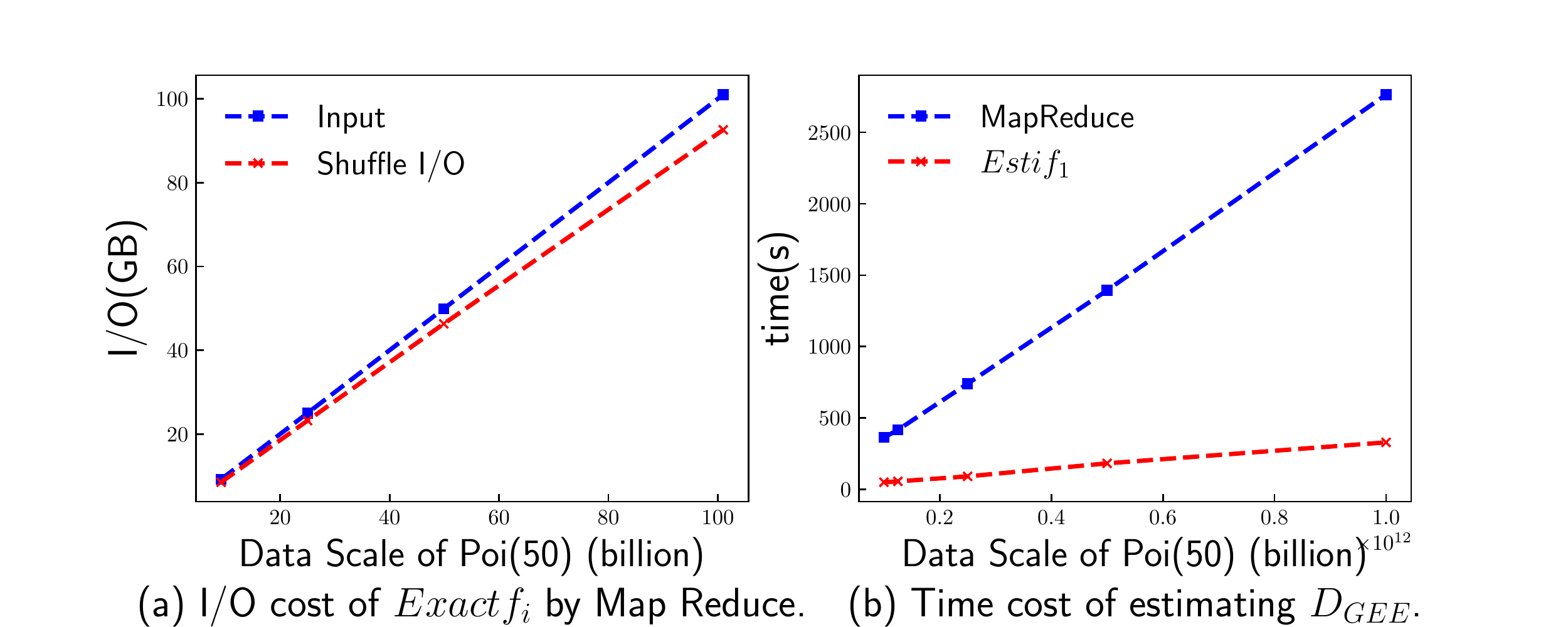}
	\vspace{-1em}
	\caption{I/O cost and Time Cost of $\hat{D}_{GEE}$}
		\vspace{-1em}
	\label{fig:spark}

\end{figure}

\header{\bf Results.} The experiment results are shown in Figure~\ref{fig:spark} and Table~\ref{tab:exp2.3}. We roughly observe the I/O cost of $Exact f_i$ implemented by MapReduce on Spark's Web UI and draw Figure~\ref{fig:spark}(a). Figure~\ref{fig:spark}(a) illustrates that $Exact f_i$ implemented by MapReduce can cause a high I/O pressure for distributed systems. The blue line in Figure~\ref{fig:spark}(a) is the I/O caused by reading data. The red line in Figure~\ref{fig:spark}(a) is the shuffle I/O caused by MapReduce. With the amount of data increasing, I/O costs grow linearly. Our algorithm, $Esti f_i$ reads and processes the data in each machine's memory, so we do not have shuffle I/O. Figure~\ref{fig:spark}(b) takes $\hat{D}_{GEE}$ as an example and shows that our algorithm is almost ten times faster than the original MapReduce algorithm.

\header{\bf Sensitive Analysis. }Recall that the main idea of our algorithm is to estimate the {\em frequency of frequency} of sample. We perform our algorithm on different scales and distributions to evaluate the efficiency of estimating $f_1$. Table~\ref{tab:exp2.3} shows the relative error of $\hat{f}_1$ based on our algorithm. When data size grows, the data will be distributed into more partitions. We can still guarantee that the relative error of $f_1$ is small. We also show that our algorithm has a negligible effect on the accuracy of sampling-based estimators in Appendix~\ref{app-exp2}.

\begin{table}[t]
	\begin{small}
		\centering
		\caption{$f_1$ Estimation with Different Scale Data in Different Distribution(HyperLogLog's $b=16$)}
		\vspace{-1em}
		\scalebox{0.9}{
		\begin{tabular}{r l l | c l l}
			\toprule
			\hline
			Distribution& $N$ &$\frac{|f_1-\hat{f_1}|}{f_1}$ &
			Distribution& $N$ &$\frac{|f_1-\hat{f_1}|}{f_1}$ \\
			\hline
			Poi(200)&1e+11 & 3.94e-01 &Zipf(1.2)&1e+11&2.15e-02\\
			Poi(200)&5e+11 & 4.50e-01 &Zipf(1.2)&5e+11&3.51e-02\\
			Poi(100)&1e+11 & 1.00e-01 &Zipf(1.5)&1e+11&2.43e-03\\
			Poi(100)&5e+11 & 1.38e-01 &Zipf(1.5)&5e+11&1.04e-02\\
			Poi(50)&1e+11&7.8e-03 &Zipf(2)&1e+11&5.38e-03\\
			Poi(50)&5e+11&1.14e-02 &Zipf(2)&5e+11&6.48e-03\\
			Poi(50)&1e+12&4.89e-03 &Zipf(2)&1e+12&4.13e-03\\
			\bottomrule
		\end{tabular}
		}
		\label{tab:exp2.3}
	\vspace{-2em}
	\end{small}
\end{table}

\section{Conclusion}

In this paper, we focus on a fundamental problem: extending various classic NDV estimators to distributed models with low communication costs. We propose a computation framework to estimate the number of the unique values of data in the distributed model. We also provide theoretical analysis for the communication cost of $f_1$ and GEE's estimators. Experiments on different synthetic datasets and the distributed environment demonstrate the efficiency of our algorithm. Based on the measurement of simulated and real-world experiments, we show that our method reduces communication costs. In future work, we intend to extend our algorithm to incorporate other estimations, such as entropy, by utilizing various types of sketches in the distributed environment.

\section{ACKNOWLEDGEMENTS}
This research was supported in part by Beijing Natural Science Foundation (No. 4222028), by National Natural Science Foundation of China (No. 61972401, No. 61932001), by the major key project of PCL (PCL2021A12), by Beijing Outstanding Young Scientist Program No. BJJWZYJH012019100020098 and by Alibaba Group through Alibaba Innovative Research Program. We also wish to acknowledge the support provided by Intelligent Social Governance Interdisciplinary Platform, Major Innovation and Planning Interdisciplinary Platform for the “Double-First Class” Initiative, Public Policy and Decision-making Research Lab, Public Computing Cloud, Renmin University of China.


\bibliographystyle{ACM-Reference-Format}

\bibliography{ndv}


\appendix

\section{Proof Of Theorem}

\subsection{Proof Of Theorem\ref{thm:1}}


\begin{proof}
We begin with the problem of set intersection. Given two sets, $A=(a_1,a_2,\ldots, a_m)$ and $B=(b_1,b_2,\ldots,b_m)$, the size of $A\cap B$, represented by $|A\cap B|$ must be calculated. The sets can be expressed as frequency vectors. Assume the initial configuration, where $X$ and $Y$ denote the frequency vectors of $A$ and $B$, respectively. If element $a$ is a member of set $A$, we have key $a$ and the associated value $1$ in frequency vector $X$. If element $a$ does not belong to set $A$, nothing of $a$ can be stored in $X$. If an element is a member of $A\cap B$, it will appear more than once in the final merging frequency vector. We convert the size of two sets' intersection, $A\cap B$, into calculating $d-f_1$ of $X+Y$. Thus, the set intersection problem can reduce to the problem of calculating the merge frequency dictionaries of two machines, i.e., calculating the $f_1$ and $d$ of $X+Y$. According to~\cite{tamm1995deterministic}, the communication complexity of $C(m_n)$ of the cardinality of the set intersection will be determined up to one bit:
$$
n+\lceil \log_2(n+1) \rceil -1 \leq C(m_n) \leq n + \lceil \log_2(n+1)\rceil .
$$
The set intersection has a communication complexity of $\Omega(m)$. Now, we demonstrate that estimating $f_1$ with relative error is equivalent to using $f_1$ to detect if an element is in the intersection of two sets. The essence of the set intersection problem is to determine whether an element takes a value of 0 or 1 in the intersection results. If we solve the set intersection problem using the relative-error estimation of $f_1$, the element takes either 0 or $v\in(\varepsilon,\frac{1}{\varepsilon})$. Thus, even if we estimate $f_1$ with relative error, we can still solve the set intersection problem. Assume that there is an algorithm that can accurately predict $f_1(X+Y)$ with relative error but has a communication complexity of $o(m)$. Then we can use this algorithm to solve the set intersection problem with communication complexity, a contradiction. As a result, the communication complexity associated with estimating $f_1(X+Y)$ with relative error is $\Omega(m)$
\end{proof}

\subsection{Proof Of Theorem\ref{the-1}}


\begin{proof}
Calculating the elements that exist only once in a given machine $A$ is expressed as $f_1^{(A)} = \big|S_{f_1^A} \cup S_{d^{\lnot A}} \big| - \big| S_{d^{\lnot A}} \big| $. The error of $f_1^{(A)} $ is equal to the sum of the errors of $S_{f_1^A} \cup S_{d^{\lnot A}}$ and $S_{d^{\lnot A}} $, which are constrained by $\varepsilon_d d$. With Assumption~\ref{ass-1}, we obtain a relative error estimation of $f_1$. Since we have $k$ machines and calculating $f_1$ requires $\ell_0$ sketches about $S_{f_1^{(X)}}$ and $S_{d^{\lnot X}}$, which cost $O(k)$ HyperLogLog($\delta,n$). One HyperLogLog($\delta,n$) costs 
\begin{equation}\label{hugeerror}
O(1/\varepsilon_d^2\log(1/\delta)\log\log n )
\end{equation}
bits. Calculating $f_1$ takes $2k$ HyperLogLog, so the communication cost is $$
O(\frac{k}{\varepsilon_d^2}\log(1/\delta)\log\log n) bits.
$$

Algorithms~\ref{alg-Pre} needs extra $O(k)$ HyperLogLog and $O(k)$ merge operations. Pre-Merged Sketches have $\log k$ levels so Algorithms~\ref{alg-Estif1} need $O(k\log k)$ merge operations in total.  
\end{proof}

\subsection{Proof Of Theorem\ref{the-3}}
\begin{proof}
Following the proof of~\cite{charikar2000towards}, the original expected value of estimator GEE is within a factor of $e\sqrt{n/r}(1+o(1))$ of correct answer. According to Theorem~\ref{the-2}, the relative error of $d$ and $f_1$ is less than $\epsilon$. $\hat{D}_{GEE}$ has a positive linear relationship with $f_1$ and $d$. So $f_1$ and d will contribute a factor $\varepsilon$ to the correct answer. Following the proof of~\cite{charikar2000towards}, the expected ratio error of $\hat{D}_{GEE}$ based on our algorithm will be $O(\varepsilon\sqrt{n/r})$.
\end{proof}

\section{Other Related Works}

\subsection{Other Related Sketch}
\header{\bf $\ell_p$ norms estimation. }Considering the norms of frequency, $\ell_0$ represent the distinct values of data. $\ell_p$ norms estimation is also a widely studied problem in stream models. Equation~\eqref{eq:fan} gives the relationship between $\ell_p$-norms with {\em frequency of frequency}. We just use $\ell_2$ norms of $f$ to estimate $\hat{D}_{CL1}$. If the estimators become more complicated than $\hat{D}_{CL1}$, we may use others $\ell_p$-norms to approximate. In other words, it is possible to derive an estimator by $\ell_p$ Sketch. $\ell_p$ sketch gives an estimation of the $\ell_p$ norm of vector $v$ in the communication complexity model. $\ell_p$ sketch utilizes a stable distribution proposed by Chambers et al.~\cite{chambers1976method} to estimate $\ell_p$ norms with a relative error. Limited by the updating time of the original $\ell_p$ sketch, Li~\cite{li2007very} proposes a faster updating method with a sparse random projection.

\subsection{Other Estimates }\label{A-Esti}
In this section, we will give some detail for complicated estimators. We also give the reason why our frameworks can handle different estimators. 

\header{\bf Chao Lee's Estimator. }
We have introduced the final expression of the first Chao Lee's Estimator before. In this part, we will give more details of Chao Lee's estimators.

We begin with some definitions of statistics. Sample coverage C is defined as the fraction of classes in the population that appears in the sample:
\begin{align*}
C=\sum_{j:n_j>0} \frac{N_j}{N}. 
\end{align*}
According to Turing et al.~\cite{good1953population}, $\hat{C}=1-f_1/n$ is used for sample coverage. To deal with the skew in the data, Chao and Lee~\cite{chao1992estimating} combine this coverage estimator with a correction term and obtain the estimator
\begin{equation}\label{eq:cl}
\hat{D}_{CL}=\frac{d}{\hat{C}}+\frac{n(1-\hat{C})}{\hat{C}}\hat{\gamma}^2,
\end{equation}
where $\hat{\gamma}^2$ is an estimator of $\gamma^2$, the squared coefficient of variation of the frequencies as follow.
\begin{align*}
\gamma^2=\frac{(1/D)\sum_{j=1}^D (N_j-\bar{N})^2 }{\bar{N}^2}. 
\end{align*}
According to~\cite{chao1992estimating}, there are following estimator of $\gamma^2$:
\begin{align*}
\hat{\gamma^2}=\max\left\{  \frac{\hat{D}_1 \sum i(i-1) f_i}{n^2-n-1},0 \right\},
\end{align*}
where $\hat{D}_1$ is an initial estimator $\hat{D}_1=d/\hat{C}$. From the above estimator of $\gamma^2$ and~\eqref{eq:cl},~\cite{chao1992estimating} constructs the following estimators:
\begin{equation}\label{eq:cl11}
\hat{D}_{CL1}=\frac{d}{\hat{C}}+\frac{n(1-\hat{C})}{\hat{C}}\hat{\gamma^2}.
\end{equation}
Chao and Lee~\cite{chao1992estimating} also introduce some improved estimators based on Equation~\eqref{eq:cl11} and other assumptions, but for these Chao Lee's estimators, $\|f\|_2$ is enough to estimate.

\header{\bf Jackknife Estimator. }Haas et al.~\cite{haas1998estimating} propose a family of estimators, with the generalized jackknife approach~\cite{gray1972generalized}. It is of the form
\begin{equation}\label{eq:JE}
\hat{D}=d+K\frac{f_1}{n}.
\end{equation}
Different approximations for $K$ results in other estimators for $D$. At first, ~\cite{haas1998estimating} obtain the first-order estimator:
\begin{equation}\label{eq:uj1}
\hat{D}_{uj1}=\left( 1-\frac{(1-q)f_1}{n}  \right)^{-1} d. 
\end{equation}

The second-order estimator $\hat{D}_{uj2}$ is derived with the approximation of $\gamma^2$. Following but different form Chao and Lee~\cite{chao1992estimating},~\cite{haas1998estimating} uses a natural method-of-moments estimator $\hat{\gamma}^2_{Hass}(D)$ of $\gamma^2$ as follows.

\begin{equation}
\hat{\gamma}^2_{Haas}(\hat{D})=\max\left(0,\frac{\hat{D}}{n^2}\sum_{i=1}^n i(i-1)f_i +\frac{\hat{D}}{N}-1 \right). \label{eq:hass}
\end{equation}
With Taylor approximations for $K$ and~\eqref{eq:hass},~\cite{haas1998estimating} obtain the second-order jackknife estimator,
\begin{equation}\label{eq:uj2}
\hat{D}_{uj2}=\left( 1-\frac{(1-q)f_1}{n}  \right)^{-1} \left(d-\frac{f_1(1-q)\ln(1-q) \hat{\gamma}^2(\hat{D}_{uj1}) }{q} \right). 
\end{equation}
By replacing the expression $f_1/n$ with approximation to $E[f_1]/n$,~\cite{haas1998estimating} obtain a smoothed second-order jackknife estimator
\begin{equation}\label{eq:sj2}
\hat{D}_{sj2}=(1-(1-q)^{\tilde{N}})^{-1}(d-(1-q)^{\tilde{N}}\ln(1-q)N\hat{\gamma^2}(\hat{D}_{uj1}),
\end{equation}
where $\tilde{N}$ is an estimate of the average class size and is set to $N/\hat{D}_{uj1}$

Although Jackknife estimator is very complicated, $\|f\|_2$ will still be enough to estimate. Since $\hat{D}$ has a more concise form, we no longer put Jackknife estimator in the experimental comparison. 

\header{\bf Shlosser's Estimator. }Shlosser's estimator also conforms to the model~\eqref{eq:JE} with parameter
\begin{align*}
K=K_{Sh}= n \frac{f_1\sum_i (1-q)^i f_i }{ \sum_i iq(1-q)^{i-1} f_i }. 
\end{align*}
Replacing $K_{Sh}$ with approximation form,~\cite{haas1998estimating} obtain the following two estimators:
\begin{equation}\label{eq:sh2}
\hat{D}_{Sh2}=d+f_1\left(\frac{q(1+q)^{\tilde{N}-1}}{(1+q)^{\tilde{N}}-1}\right)\left( \frac{ \sum_{i=1}^n (1-q)^i f_i }{\sum_{i=1}^n iq(1-q)^{i-1}f_i}\right),
\end{equation}
where $\tilde{N}$ is an estimate of the average class size and is set to $N/\hat{D}_{uj1}$

When calculating $\hat{D}_{Sh2}$ for large $N$, it will result in floating point errors. Following~\cite{deolalikar2016extensive}, we can use $q/(1+q)$ to approximate the term in the first parentheses in~\eqref{eq:sh2}. For more complicated Shlosser's estimators, we still can resample to approximate, which also proves the applicability of our algorithm.


\section{Additional Experimental Results}
\subsection{Simulated Experiments}\label{app-exp}
\begin{figure}[t] 
	\centering
	\includegraphics[width=85mm]{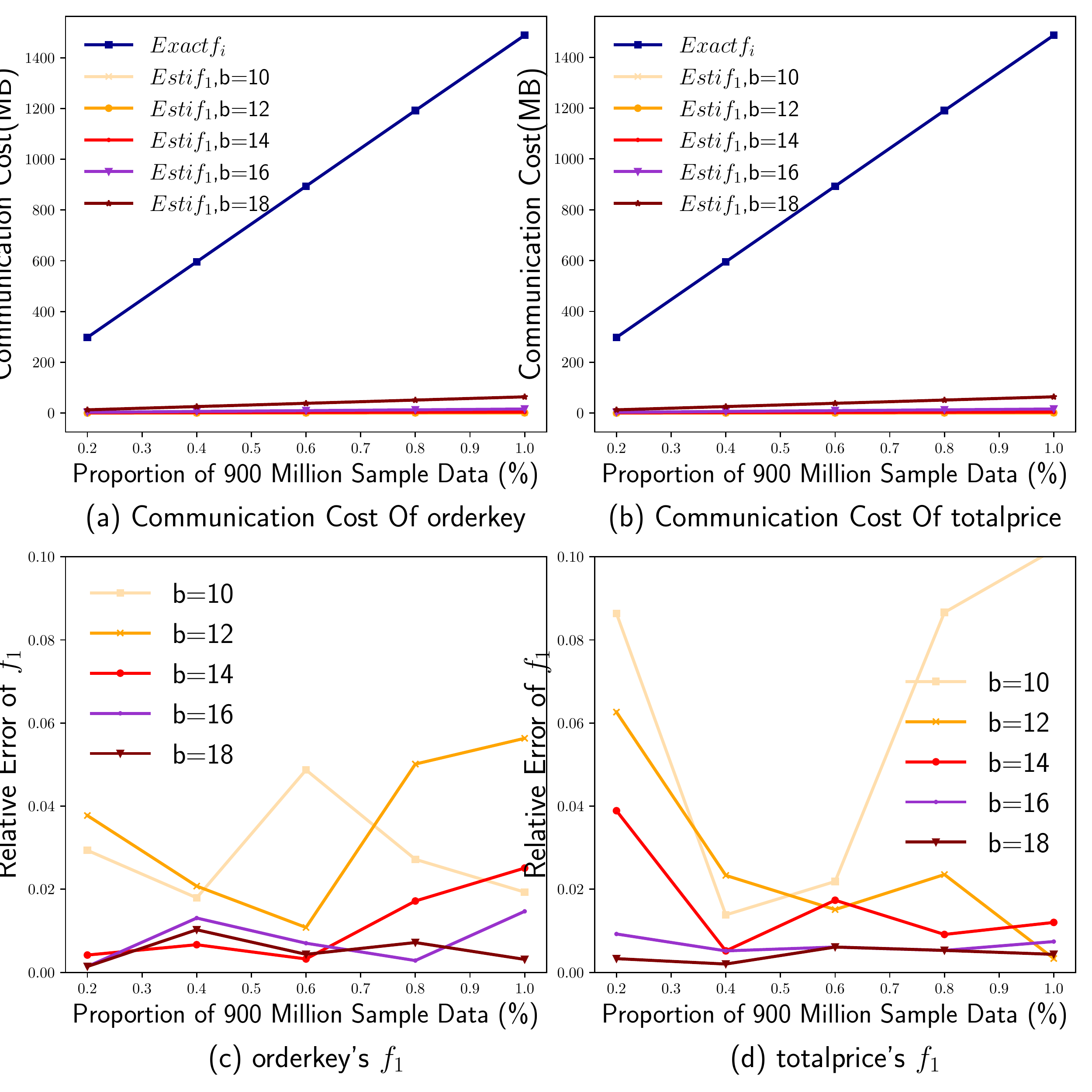}
	\caption{Communication Cost of Different Methods with Different Parameter. $\lambda$ is the parameter of Poisson distribution; s is the parameter of Zipfian distribution; b is the parameter of HyperLogLog.}
	\label{fig:order}
	\vspace{-1em}
\end{figure}

We also evaluate the $Exact f_i$ and $Esti f_i$ on orderkey and totalprice. Figure~\ref{fig:order} shows the communication costs and the relative error of two data. The conclusion is the same as revenue's. The communication cost of $Esti f_i$ is lower than $Exact f_i$'s on both synthetic data and real-world data. When the parameter $b$ of HyperLogLog is small than $12$, we can have a relative error estimation for $f_1$.

\subsection{Experiments on Spark}\label{app-exp2}

We also evaluate the different estimators implemented by our method on Spark.  Table~\ref{tab:exp2.2} shows the performance of different estimators. The conclusion is the same as the simulated experiments. Our method indeed has a negligible effect on the accuracy of sampling-based estimators. 

\begin{table}[t]
	\begin{small}
		\centering
		\caption{Relative Error Of Our Method. Values without parentheses denote the relative error about estimators and values with parentheses denote the relative error about true $D$ }
		\vspace{-1em}
		\scalebox{0.9}{
		\begin{tabular}{l c c c c  }
			\toprule
			\hline
		Parameter  & $\hat{D}_{GEE}(\hat{D})$ & $\hat{D}_{Chao2}(\hat{D})$&$\hat{D}_{Sh}(\hat{D})$&$\hat{D}_{CL1}(\hat{D})$\\
			\hline
		$N=1e+11, \lambda=50$&0.011(2.1)&0.12(0.12)&0.016(23)& 0.45(0.024)\\
		$N=5e+11, \lambda=50$&0.012(2.1)&0.067(0.067)&0.018(23)&0.42(0.026)\\
		$N=1e+11, \lambda=100$&0.022(2.9)&0.13(0.13)&0.080(25)&0.38(0.020)\\
		$N=5e+11, \lambda=100$&0.011(3.0)&0.10(0.10)&0.14(26)&0.36(0.013)\\
		$N=1e+11, \lambda=200$&0.027(2.2)&0.083(0.083)&0.19(14)&0.25(0.010)\\
		$N=5e+11, \lambda=200$&0.015(2.4)&0.65(0.065)&0.27(15)&0.23(0.011)\\
		$N=1e+11, s=1.2$& 0.011(0.78)&0.40(0.88)&0.19(1.1)&0.92(0.92) \\
		$N=5e+11, s=1.2$ &0.0051(0.77)&0.36(0.87)&0.11(0.95)&0.92(0.92)\\
		$N=1e+11, s=1.5$ &0.011(0.83)&0.35(0.81)&0.21(0.56) &0.84(0.86)\\
		$N=5e+11, s=1.5$ &0.0017(0.83)&0.27(0.79)&0.035(0.65)&0.82(0.84)\\
		$N=1e+11, s=2$ &0.0053(0.87)&0.22(0.45)&0.0097(0.31)&0.68(0.54)\\
		$N=5e+11, s=2$&0.0058(0.87)&0.46(0.62)&0.0061(0.31)&0.69(0.56)\\
			\bottomrule
		\end{tabular}}
		\label{tab:exp2.2}
		\vspace{-1em}
	\end{small}
\end{table}

\end{document}